\documentclass[a4paper,aps,pra,twocolumn,showpacs]{revtex4}             
\usepackage{bbm}                                                      
\usepackage{amsmath}                                                  
\def\RR{\ensuremath{\mathbbm{R}}}                                     
\def\CC{\ensuremath{\mathbbm{C}}}                                     
\def\id{\ensuremath{\mathbbm{1}}}                                     
   
\def\LZRn{\ensuremath{\mathrm{L}^{\!\rule[-0.5ex]{0mm}{0mm}2}(\RR^n)}}
\def\cB{{\cal B}}                                                     
                                                     
\def\cE{{\cal E}}                                                     
\def\cG{{\cal G}}                                                     
\def\cH{{\cal H}}

\def\cBH{{\cal B}(\cH)}                                               
\def\cGH{{\cal Q}(\cH)}                                               
\def\tr{\mathrm{tr}}                                                  
\def\ket#1{\left| #1\right>}                                          
\def\bra#1{\left< #1\right|}                                          
\renewcommand{\Re}{\mathrm{Re}}                                       
\renewcommand{\Im}{\mathrm{Im}}                                       
\newcommand{\Eqref}[1]{Eq.~(\ref{#1})}                                
\newcommand{\Eqsref}[1]{Eqs.~(\ref{#1})}                             
\begin{document}                                                      
\title{The characterization of Gaussian operations and Distillation of
Gaussian States}        
                                                                      
\author{G\'eza Giedke and J. Ignacio Cirac}                           
                                                                      
\affiliation{Max-Planck--Institut f\"ur Quantenoptik,                 
Hans-Kopfermann-Strasse, D-85748 Garching, Germany}                  
                                                                      
\begin{abstract}                                                      
We characterize the class of all physical operations that transform
Gaussian states to Gaussian states. We show that this class coincides
with that of all operations which can be performed on Gaussian states
using linear optical elements and homodyne measurements. For bipartite
systems we characterize the processes which can be implemented by
local operations and classical communication, as well as those that
can be implemented using positive partial transpose preserving
maps. As an application, we show that Gaussian states cannot be
distilled by local Gaussian operations and classical communication. We
also define and characterize positive (but not completely positive)
Gaussian maps.
\end{abstract}                                                        
                                                                      
                                                                      
\maketitle                                                            
                                                                      
\section{Introduction}                                                                      
Many applications in the field of quantum information require the
ability of preparing general states and of performing arbitrary
transformations with them. However, there are physical systems where
the set of states that can be generated as well as the transformations
that can be implemented are very restricted. For example, in quantum
optical systems linear transformations involving beam splitters, phase
plates, homodyne measurements and polarizers are readily implemented
whereas more general ones can only be performed with a low
efficiency. Moreover, with these tools and a squeezer one can generate
only a small class of states, the so--called Gaussian states. Despite
this fact, in these systems a surprising richness of quantum
information protocols has been found within the realm of linear
optics: entanglement generation \cite{Reid}, teleportation
\cite{CVQTelTh}, key distribution \cite{CVQKD}, quantum error
correction \cite{CVQEC}, cloning \cite{CVQClon}, some of which have
already been implemented
\cite{allexperiments}. For the moment it is not known how (or whether)
important operations such as entanglement distillation or a (useful)
depolarization of continuous variable states can be implemented with
linear optics
\cite{Opa99,Par99,Duan99b,DCVS,Thesis,LOG}.        
This raises the question which transformations can in general be       
realized by the concatenation of such operations, i.e. the tools      
currently available in the lab.                                       
                                                                      
A partial answer to this question is contained in a mathematical      
paper written in the seventies \cite{DVV}. There a subclass of        
operations that transform Gaussian states into Gaussian states        
have been mathematically characterized, namely those which are        
\emph{trace-preserving}. The relation of these maps to some of
the experimentally feasible operations has been discussed by Eisert
and Plenio in \cite{LOG}, and used by these authors to derive a
criterion for the interconvertibility of two-mode Gaussian states
under feasible local transformations.  
Unfortunately, the set of operations considered in Ref. \cite{DVV}    
does not include measurements (which are not trace-preserving).       
However, one of the main strengths of linear optics is the highly     
efficient measurement of the quadrature observables $X$ and $P$       
which homodyne detection affords. Furthermore, measurements           
followed by classical communication have been seen to be an           
essential ingredient in certain basic quantum information             
protocols such as quantum teleportation \cite{QBQTelTh} or            
entanglement distillation \cite{Ben96}. Thus it is important to     
find a mathematical formulation of the physical actions that can      
be applied to Gaussian states in which also not trace-preserving      
operations (measurements) are included \cite{FN0}. 

In this work we give a        
full answer to this problem by providing a simple description of      
all operations of this sort. As an application we discuss the question
of distillation of Gaussian states  by local Gaussian             
operations and classical communication (LOGCC). In this context, 
it was recently shown \cite{nogo} that the
entanglement of a symmetric two-mode Gaussian state of two parties
cannot be increased with the help of another copy of that state and a
homodyne measurement. Here we give a proof that distillation is not
possible for an arbitrary number of modes per site, general Gaussian
states and general Gaussian operations. 

This paper is organized as follows. In Section \ref{SecGStates} we fix
our notation for Gaussian states, setting the stage for the results
given in the following sections. Section \ref{main} contains the main
results of this paper: we give a characterization of all completely
positive maps that transform Gaussian states into Gaussian states
(Gaussian operations). We show that they can all be implemented with
the currently available means. We derive a simple, compact form of
these maps. In Section \ref{Secbipart} we consider Gaussian operations
on bipartite systems; we classify them with respect to their locality
and separability properties. As an application of the methods
introduced before, we show that Gaussian states cannot be distilled by
using Gaussian LOCC. In Section \ref{Secposmaps} we describe positive Gaussian
maps and characterize them completely. Appendix \ref{V} contains
some material on a new entanglement measure for Gaussian states
introduced and used in Sec. \ref{Secbipart}.  
%Finally, in the appendix 
%we describe positive Gaussian
%maps and characterize them completely.
                                                                      
%%%%%%%%%%%%%%%%%%%%%%%%%%%%%%%%%%%%%%%%%%%% 
\section{Gaussian States}\label{SecGStates}
We consider the Hilbert space of $n$ harmonic oscillators
$\cH=\LZRn$. A Gaussian state is described by a density
operator $\rho$ whose characteristic function \cite{Petz}
$\chi_{\rho}(x):=\tr[\rho W(x)]$ is a Gaussian function in
$x\in\RR^{2n}$ 
(or, equivalently, by a Gaussian Wigner function, which is related to
$\chi$ by a symplectic Fourier transformation). 
The operators
\begin{equation}\label{Weyl}                                          
W(x) = \exp\left[-ix^TR  \right],                                     
\end{equation}                                                        
are the \emph{Weyl operators} (displacement operators) and $R =
(X_1,P_1,X_2,\dots,P_n)$, with $[X_k,P_l]=i\delta_{kl}$.  Since the
density operator is bounded, we can write without loss of generality
\cite{Perelomov}
\begin{equation}\label{GaussOp}                                       
\rho = \pi^{-n}\int_{\RR^{2n}}\!\!dx e^{-\frac{1}{4}x^T\gamma x +       
id^Tx}W(x),                                                            
\end{equation}
where we have used $\tr[W(x)]=\pi^ n\delta(x)$ \cite{Perelomov} to
normalize $\rho$. Occasionally, we will denote $\rho$ as in
\Eqref{GaussOp} by $\rho_{\gamma,d}$. 
The matrix $\gamma=\gamma^T\ge iJ_n$ is a $2n\times 2n$ real matrix     
called \emph{correlation matrix} (CM) and $d$ is a $2n$ real vector          
called \emph{displacement}. These two quantities fully characterize the      
Gaussian state $\rho$. The \emph{symplectic matrix} $J_n$ is          
\begin{equation}\label{symplect}                                      
 J_n = \bigoplus_{k=1}^n J_1, \quad J_1 =                             
\left(\begin{array}{cc}0&-1\\1&0                                      
\end{array} \right),                                                  
\end{equation}                                                        
(We will omit the index whenever there is no risk of confusion.)      
Note that the CM usually contains all the interesting information     
about the properties of the state which are useful for quantum        
information in general; in particular, the entanglement properties of
a Gaussian state is solely determined by its CM. Thus, in some parts
of this paper in order to simplify the notation we will omit the
displacement from our 
discussions when it does not play a relevant role.

Of course, Gaussian states can be also defined for composite systems,
e.g. those whose density operators are in ${\cal B}({\cal H})\otimes
{\cal B}({\cal H})$. An important example of a Gaussian state is the
maximally entangled state $\Phi$ \cite{footnote}. The state 
$\Phi$ is the limit $r\to\infty$ of Gaussian states
($n$ identical two--mode squeezed states) with CM
\begin{equation}\label{maxentCM}                                      
\gamma(r) = \left( \begin{array}{cc} A_r&C_r\\C_r&A_r                 
\end{array} \right)                                                   
\end{equation}                                                        
where $A_r = \cosh r\id $ and $C_r=\sinh r \Lambda$ are               
$2n\times2n$ square matrices, and 
\begin{equation}                                                      
\Lambda = \text{diag}(1,-1,1,-1,\dots,-1). 
\end{equation}                                                                      
The density operator of $\Phi$ is a projector on the improper state
vector $\ket{\Phi}_{12}\propto\sum_{k\geq0}\ket{k}_1\ket{k}_2$.  

%%%%%%%%%%%%%%%%%%%%%%%%%%%%%%%%%%%%%%%%%%%% 
\section{Gaussian Operations}\label{main}
Physical actions are mathematically characterized in terms of
completely positive (cp) maps acting on the corresponding density
operators. The best way of characterizing them is using the
isomorphism between cp maps (physical actions) and positive operators
(unnormalized states) \cite{isomorphism}.
                 
\subsection{General form of Gaussian operations}
Given a cp map ${\cal E}$ acting on bounded operators ${\cal B}({\cal
H})$, we define the positive operator $E\in {\cal B}({\cal H})\otimes
{\cal B}({\cal H})$ as follows
\begin{equation}\label{EQ}                                            
E_{12}=({\cal E}\otimes\id)(|\Phi\rangle_{12}\langle\Phi|).
\end{equation}
This equation has a direct physical meaning. It tells us that given
${\cal E}$, we can always obtain the state $E$ by preparing a
maximally entangled state and acting with the map on the second
subsystem. Conversely, given the state $E\in {\cal B}({\cal
H})\otimes {\cal B}({\cal H})$ and a state $\rho\in {\cal
B}({\cal H})$, if we measure the second subsystem of $E$ and $\rho$ in
the ``Bell basis'', i.e., an orthonormal basis of maximally entangled
states containing $\ket{\Phi}$, and obtain the result corresponding to
the state 
$|\Phi\rangle$, then the resulting state is ${\cal E}(\rho)$. Thus,
given the state $E$ we can always implement (probabilistically) the
map ${\cal E}$ provided we can perform Bell measurements. This can be
viewed as ``teleporting $\rho$ through the gate $\cE$''
\cite{telepQC}. In formulas, we have
\begin{equation}\label{GviaTelep}
{\cal E}(\rho)\propto \tr_2[E_{12}^{T_2}\rho_2] =
\tr_{23}(E_{12}\rho_3\ket{\Phi}_{23}\bra{\Phi})
\end{equation}                                                        
Thus, given a cp map we can generate the corresponding state and
given the state we can physically implement $\cE$ (probabilistically)
\cite{FN_improper}. 
                                                                      
Now we define a Gaussian completely positive (g-cp) map, ${\cal G}$ by
the properties that both $\cG$ and $\id\otimes\cG$ map Gaussian states
to Gaussian states. With the help of the isomorphism it is
straight forward to characterize them. First, we use the fact that the
state $|\Phi\rangle$ appearing in (\ref{EQ}) is Gaussian, so that the
corresponding operator $G$ must be Gaussian \cite{useisomorphG}. We write it as
\begin{equation}\label{GaussOp2}           
G = \int_{\RR^{4n}}\!\!dx e^{-\frac{1}{4}x^T\Gamma x +          
iD^Tx-C}W(x).                                                           
\end{equation}
Clearly $G\geq0$ iff $\Gamma, D, C$ are real and $\Gamma\geq iJ$; that
is, $\cG$ is a g-cp map if the Gaussian operator isomorphic to $\cG$ is
described by a proper CM $\Gamma$ and, conversely, to each such operator
corresponds a g-cp map $\cG$. Since all Gaussian states can be
generated (e.g., from the vacuum state) by unitary Gaussian operations
and discarding subsystems \cite{Simo94} this shows that all Gaussian
operations can be implemented by these means plus Bell measurements
(homodyne detection).  

Now, we determine the action of the map ${\cal G}$ on a general
Gaussian state $\rho$ in terms of $G$. Apart from normalization, we
have
\begin{equation}\label{GMap}
\cG:\rho_{\gamma,d} \mapsto \rho_{\gamma',d'},
\end{equation}
and we find for $\gamma',d'$:
\begin{subequations}\label{gammap_gamma}
\begin{eqnarray}                                                      
\gamma' &=&                                                           
\tilde\Gamma_1-\tilde\Gamma_{12}\frac{1}{\tilde\Gamma_2+\gamma}     
\tilde\Gamma_{12}^T,\\                                                  
d' &=& D_1+\tilde\Gamma_{12}\frac{1}{\tilde\Gamma_2+\gamma}(D_2+d),
\end{eqnarray}                          
\end{subequations}                              
where we have denoted
\begin{equation}\label{capGamma}                                      
\Gamma = \left( \begin{array}{cc} \Gamma_1&\Gamma_{12}\\              
\Gamma_{12}^T&\Gamma_2                                                
\end{array}\right)\!,\, 
D = \left( \begin{array}{c} D_1\\ D_2\end{array} \right),   
\end{equation}                                                        
and                                                                   
\begin{equation}\label{restriction}                                   
\tilde\Gamma =
(\id\oplus\Lambda)\Gamma(\id\oplus\Lambda).
\end{equation}                                                        
Thus, we have that all g-cp maps on $\cBH$ are characterized by a
correlation matrix $\Gamma\geq iJ_{2n}$ and a displacement vector $D$.
                                                                      
In order to derive Eqs. (\ref{gammap_gamma}), we just use
\Eqref{GviaTelep}, replacing $E_{12}$ by $G$ and $\rho_2$ by a
Gaussian state $\rho_{\gamma,d}$ as in \Eqref{GaussOp}. In evaluating
the trace we use the commutation relation $W(x)W(y)=e^{i/2x^TJy}W(x+y)$
\cite{Petz} and $\tr[W(x)] \propto
\delta(x)$ to obtain
\begin{equation}\label{Gdensmatop}
\rho_{\gamma',d'} \propto
\tr_2(G^{T_2}\rho_{\gamma,d}), 
\end{equation}
with $\gamma',d'$ as in Eqs. (\ref{gammap_gamma}). 

\subsection{Examples}\label{examples}
How to interpret the operation described by $(\Gamma, D)$? 
To better understand what actions $(\Gamma,D)$ describe, we now
briefly discuss how the familiar Gaussian operations are contained in
our formalism. To do this, we apply these operations to the first
subsystem of the maximally entangled Gaussian state $\Phi$ with CM
$\Gamma=\lim_{r\to\infty}\gamma(r)$, with $\gamma(r)$ as in
\Eqref{maxentCM}.   

Obviously, the identity operation corresponds to the maximally
entangled state $\Phi$, i.e., to $\Gamma=\lim_r\gamma(r), D=0$. 
Now, performing a displacement operation on $\Phi$ leaves
the CM unchanged, but produces a displacement $D=(D_1,0)$. 
Now we turn to the trace-preserving g-cp maps considered in
\cite{DVV}. These describe all actions that can be performed on $\rho$
by first adding ancillary systems in Gaussian states, then performing
unitary Gaussian transformations on the whole system, and finally
discarding the ancillas. On the level of CMs these operations 
were shown to be described by $\gamma\mapsto M^T \gamma M+N$. 
A Weyl operator $W(x)$ is mapped to $(\det M)^{-1}e^{-1/4
x^T(M^{-1})^TNM^{-1}x}W(M^{-1}x)$ by these 
operations. It then follows that the Gaussian operator that
corresponds to this operation has the CM 
\[
\Gamma = \lim_{r\to\infty} \left( \begin{array}{cc} 
M^T A_r M + N & M^TC_r\\
C_r M& A_r
\end{array} \right).
\]
Using formulas (\ref{gammap_gamma}) gives
$\gamma' = \lim_r M^T A_r M + N - M^TC_r\Lambda(\Lambda A_r\Lambda +
\gamma)^{-1}\Lambda C_rM$. For $r\to\infty$ we have $(\Lambda A_r\Lambda +
\gamma)^{-1}\to A_r^{-1} - A_r^{-1}\gamma A_r^{-1}+o(\cosh r)^{-3}$
which yields the desired result \cite{FN2}. 

Finally, we consider an example of Gaussian measurements. The typical
measurement is homodyne detection, which realizes the von Neumann
measurement of the operator $X$. It has been shown before
\cite{Leonhardt} that with the use of an ancillary system and a beam
splitter, homodyne measurements may be used to realize the generalized
measurement corresponding to the positive-operator-valued measure
(POVM) $\{\ket{\alpha}\bra{\alpha}, 
\alpha\in\CC\}$, where $\ket{\alpha}$ is a coherent state, i.e. in the
language of CMs a state with CM $\gamma=\id$ and displacement
$d=(\Re\alpha,\Im\alpha)$. Since every other pure Gaussian state can
be obtained from $\ket{\alpha}$ by Gaussian unitaries, this implies
that \emph{all} POVMs of the form $\{\ket{\gamma,d}\bra{\gamma,d} : 
d\in\RR^{2n}\}$ can be performed with homodyne detection and suitable
pre-processing.

To see which CM $\Gamma$ corresponds to the measurement of
$\ket{\gamma,d}\bra{\gamma,d}$ we apply it to $\Phi$. In order to get an
interesting result, we perform only a partial measurement, i.e. we
consider an $(n+m)$-mode system and measure only the last $m$ modes. 
The corresponding operator is 
\[E = \bra{\gamma,d}\Phi\ket{\gamma,d},
\]
where $\ket{\gamma,d}$ describes a pure Gaussian state of $m$ modes.
Expressing $\Phi$ in terms of Weyl operators, replacing
$\bra{\gamma,d}W(x)\ket{\gamma,d}$ by $e^{-1/4 x^T\gamma x-id^Tx}$ and
integrating over the modes measured we obtain a CM 
\begin{equation}
\Gamma = \left( \begin{array}{ccc}
A_r&C_r&0\\
C_r& A_r & 0\\
0 &0& \gamma^{-1}
\end{array} \right)
\end{equation}
and a displacement $D = (0,0,d)^T$.
Note that the first row corresponds to system ``1'', while the second
and third row refer to system ``2''. This represents a straight forward
generalization of the situation considered in Section \ref{main} to
maps which decrease the number of modes present. Evaluating
\Eqref{GviaTelep} for a $n+m$ mode Gaussian state with CM~$\gamma$ 
\[\gamma = \left( \begin{array}{cc} A&C\\ C^T&B                       
\end{array} \right),                                                  
\]
we obtain \cite{FN2}
\begin{subequations}\label{aftermeas}
\begin{equation}\label{CMaftermeas}                                   
\gamma' = A - C\frac{1}{B+\gamma_p}C^T,
\end{equation}                                                        
\begin{equation}\label{dispaftermeas}                                 
d' = \frac{1}{2}C\frac{1}{B+\gamma_p}d,
\end{equation}                           
\end{subequations}
which corresponds to the change in CM derived in \cite{nogo} for projections
into pure Gaussian states. 
Homodyne detection itself represents the limiting case in which the CM
$\gamma_p$ becomes infinitely squeezed. In this limit, the inverse in
Eqs. (\ref{aftermeas}) is to be understood as the pseudo-inverse (inverse
on the range). 

In general, \emph{noise-free Gaussian operations} (unitaries and von
Neumann measurements) correspond to \emph{pure state CMs}
$\Gamma$ and noise added to the CM describing the operation directly
translates into noise added to the output state, i.e., we have that
$\Gamma'=\Gamma+P\geq\Gamma$ implies that $G_{\Gamma'}(\gamma)\geq
G_\Gamma(\gamma)\forall\gamma\geq iJ$. To see this, consider the
operation $\cG_{\Gamma'}$ and write $G$ in
\Eqref{Gdensmatop} as a mixture of states with CM $\Gamma$. This shows
that the state $\cG_{\Gamma'}(\rho_{\gamma,d})$ is a Gaussian mixture
of states $\cG_\Gamma(\rho_{\gamma,d+d_2})$ displaced by $d_1$, where
$x=(d_1,d_2)$ are distributed according to a probability distribution
proportional to $\exp(-1/4x^TP^{-1}x)$.  But since a displacement
$d_2$ of the input state does only affect the displacement of the
output state [cf. \Eqsref{gammap_gamma}], it follows that
$\cG_{\Gamma'}(\rho_{\gamma,d})$ is nothing but a Gaussian mixture of
states $\cG_\Gamma(\rho_{\gamma,d})$ displaced by some value $y$ which
is distributed according to a Gaussian distribution with covariance
depending on $P$ and $\gamma$. Thus the operation $\cG_{\Gamma'}$
could be realized (for known $\gamma$) by first performing
$\cG_\Gamma$ and then performing random displacements to add the
appropriate noise, which proves the assertion. Since displacements can
be done locally, this becomes particularly useful in the discussion of
entanglement distillation with Gaussian means below.

\subsection{Deterministic Operations}\label{detOp}
In general, the transformation \Eqref{GMap} is not
trace-preserving. This is related to the fact that we have considered
only one of the possible Bell measurements in \Eqref{GviaTelep}. The
projector $\ket{\Phi}\bra{\Phi}$ can be extended to a POVM by
considering all displacements $W(x)\ket{\Phi}\!, x\in\RR^{4n}$. Using
the second relation in \Eqref{GviaTelep} it is easy to see that if
$\cG_{\Gamma,0}(\rho) = \tr_{23}(E_{12}\rho_3\ket{\Phi}_{23}\bra{\Phi})$ then 
$\cG_{\Gamma,D}(\rho) = \tr_{23}[E_{12}\rho_3
W(-D)\ket{\Phi}_{23}\bra{\Phi}W(D)]$, i.e. $D$ can be understood as
the (continuous) output of the Bell-measurement implementing $\cG$. 

Note that $D$ has no influence on the CM of the resulting
state. Hence, provided $\gamma$ and $\Gamma$ are known, $\cG$ can be
turned into a trace-preserving operation by postprocessing:
conditional on the measurement result $(D_1,D_2)$ the corresponding
displacement $D_1+\tilde\Gamma_{12}(\tilde\Gamma_2+\gamma)^{-1}D_2$
can be undone, leading to a \emph{deterministic} transformation that
maps every Gaussian state $\rho_{\gamma,d}$ to
$\rho_{\gamma',\tilde\Gamma_{12}(\tilde\Gamma_2+\gamma)^{-1}d}$ with
certainty.  Note that this is true even if $\gamma$ is a state on a
multipartite system, since displacements can be done locally.  It is a
curious feature of Gaussian operations that even measurements do not
change the CM non-deterministically.

%%%%%%%%%%%%%%%%%%%%%%%%%%%%%%%%%%%%%%%%%%%%%%%%%%%%%%%
\section{Bipartite systems. Applications}\label{Secbipart}            
In this section we consider Gaussian maps $\cG$ on bipartite
systems. In this situation it is interesting to distinguish whether
$\cG$ can be implemented with \emph{local operations} on the
subsystems A and B (possibly enhanced by classical communication,
LOCC) or whether interaction between the system is necessary. 
Our formalism yields a very convenient form for any local 
Gaussian operations, and allows to determine the nonlocal properties
for any given Gaussian map.

As an application we use our formalism to show that entanglement
distillation is not possible with Gaussian means. This extends the
results of \cite{nogo} to any number of modes, all kinds of Gaussian
operations, and all kinds of Gaussian states. Note that, there are
other means of performing distillation which do not require Gaussian
maps as long as a Kerr non--linearity or photodetection are available
\cite{Duan99b,Opa99}. However, at the moment these protocols still
pose considerable experimental challenges and none has been
implemented to date.

\subsection{Local Gaussian maps assisted by classical                 
communication}

To determine whether $\cG$ can be implemented with \emph{local
operations} on the subsystems A and B (possibly enhanced by classical
communication, LOCC) or whether interaction between the system is
necessary we can use the the ideas of Ref. \cite{paperciracetal}
(which extend the Jamio{\l}kowski isomorphism to bipartite
systems). This allows to read off the answer to this question from the
Gaussian state $G$ isomorphic to $\cG$. If $G$ is separable, then it
can be generated by local action and classical communication. Note
that following the discussion at the end of the previous section this
implementation can be done \emph{deterministically} provided the CM of
the state on which we act is known.

If $G$ is entangled, two cases can be distinguished: is $G$ has
positive partial transpose (ppt), the corresponding map can be implemented
with a so-called ppt-preserving channel, otherwise full-fledged
quantum interaction between A and B is needed.
                                                                      
The separability criterion for Gaussian states \cite{Sim99,GSepCrit} allows 
us to decide for every given map, whether it is separable
\cite{GSepCrit}, ppt-preserving \cite{Sim99} or neither. 
Moreover, the characterization of separable CMs given in         
\cite{GBE}, namely that $\gamma$ is separable iff $\exists            
\gamma_A,\gamma_B\geq iJ$ such that $\gamma\geq\gamma_A\oplus\gamma_B$
implies that -- except for added correlated noise -- all Gaussian LOCC
operations are of product form. 
                                                                      
\subsection{Gaussian states cannot be distilled with Gaussian         
local operations and classical communication}                         

Entanglement distillation is a process in which two separate parties A
and B transform a large number of copies of a bipartite mixed
entangled state $\rho_{AB}$ (jointly written as $\rho_{AB}^{\otimes
n}$) into a state $\psi_{AB}^{(n)}$, which, as $n$ goes
to infinity approaches a pure maximally entangled state.
To this end, A and B are allowed to perform arbitrary local
operations (correlated by classical communication) on their respective
part of $\rho_{AB}^{\otimes n}$. In the following we show that such a
process is not possible when $\rho_{AB}$ is Gaussian and only Gaussian
operations are allowed. 

We consider a bipartite system composed of subsystems A and B and a
partially entangled Gaussian state with CM $\gamma_{AB}$  and want to
check if a separable Gaussian map can increase the entanglement. To
this end we define a simple function $V(\gamma)$ to quantify the
entanglement of a general bipartite CM. Let $V(\gamma_{AB})$ be the
largest value $p\leq 1$ such that $\gamma_{AB}\ge p 
(\gamma_A\oplus\gamma_B)$, for some CMs $\gamma_A$ and
$\gamma_B$. Note that for a maximally entangled Gaussian state $V=0$
[this and further properties of $V(\gamma)$ which are used in the 
following are proved in Appendix \ref{V}],
and therefore the goal of a Gaussian distillation protocol
would be to decrease $V(\gamma_{AB})$. In general one would allow an
arbitrary number copies of the state with CM $\gamma_{AB}$, i.e. a
state with the CM $\bigoplus_{k=1}^n\gamma_{AB}$. As shown in
App. \ref{V}, $V$ does not change when adding more copies of the same
state. The question then is, whether there is a local Gaussian
operation $\cG$ that produces from these states an output state
$\gamma_{AB}'$ with $V(\gamma_{AB}')<V(\gamma_{AB})$ or even allows to
reach $V\to0$ in the limit of infinitely many copies. 

In the following we show that with local Gaussian operations it is
impossible to decrease $V$ at all. Our proof makes no assumptions on
the size or type of the entangled state considered or the local
Gaussian operations performed. In particular, it covers any number of
copies of a $n\times n$-mode input state.

We consider a separable Gaussian completely positive map acting on two
systems A and B. As discussed above, the action of such a map on the
correlation matrix $\gamma_{AB}$ is completely characterized by
another correlation matrix $\Gamma$ acting on an extended space of
systems A, A', B and B'. The fact that the map is separable implies
that $\Gamma=\Gamma_{AA'}\oplus\Gamma_{BB'}+P$, where $P$ is a
positive matrix. In light of the discussion at the end of
Subsec. \ref{examples} this means that $\Gamma$ can be implemented by
first performing the (completely uncorrelated) operation corresponding
to $\Gamma_{AA'}\oplus\Gamma_{BB'}$ and then performing (classically
correlated) random displacements of the resulting state according to a
probability distribution depending on $P$ and the CM $\gamma_{AB}$ of the
input state. Since these displacements do not increase the
entanglement, we can concentrate on the effect of the product
transformation $\Gamma_{AA'}\oplus\Gamma_{BB'}$.

Now let 
\begin{equation}                                                      
 \Gamma_{AA'}=\left(\begin{array}{cc}                     
 A_1 & C_A\\ C_A^T & A_2 \\ \end{array}\right),\quad                  
 \Gamma_{BB'}=\left(\begin{array}{cc}                                 
 B_1 & C_B\\ C_B^T & B_2 \\ \end{array}\right).                       
\end{equation}                                                        
Let us denote by $\gamma_{AB}'$ the correlation matrix of A and B
after the action of the map $\Gamma_{AA'}\oplus\Gamma_{BB'}$. Then we
have $\gamma_{AB}'\ge R_A\oplus R_B$, where 
\begin{equation}\label{noinc}                  
 R_A=\tilde A_1 -\tilde C_A\frac{1}{\tilde A_2+p \gamma_A}\tilde
C_A^T\ge \tilde A_1
-\tilde C_A\frac{1}{\tilde A_2+piJ}\tilde C_A^T,
\end{equation}                                                        
and similarly for $R_B$. Now, we use that $\Gamma_{AA'}\ge 0,i(J\oplus
J)$ since $\Gamma$ is a CM and therefore $\tilde\Gamma_{AA'}\ge
p[iJ\oplus (-iJ)]$. This implies that the RHS of \Eqref{noinc} is
$\geq ip J$ and we immediately obtain that $\gamma_A'\equiv 1/pR_A$ is
a correlation matrix. From this follows that $\gamma_{AB}'\ge
p(\gamma_A'\oplus \gamma_B')$ and therefore $V(\gamma_{AB}')\ge
V(\gamma_{AB})$.
                
What does this imply for distillation? First, it proves that the
maximally entangled state cannot be approached even asymptotically
(i.e., in the limit when initially infinitely many copies of
$\gamma_{AB}$ are available). This follows directly from the fact
(cf.\ App. \ref{V}) that
$V(\gamma_{AB}\oplus\gamma_{AB})=V(\gamma_{AB})$, i.e., $V$ is
invariant when adding more copies of the same resource. So
entanglement distillation of Gaussian states with Gaussian means is
impossible. More generally, $V(\gamma_{AB})$ puts a bound on
all state transformations that can be achieved by Gaussian LOCC [and
even Gaussian 
LOCC supplemented by an unlimited amount of auxiliary entangled
Gaussian states of $V(\rho_{aux})\geq V(\gamma_{AB})$]. 

However, the result still leaves room for interesting entanglement
transformations with Gaussian means. The best thing that could happen
-- respecting the bound set by $V_{min}=V(\gamma_{AB})$ -- is to have the
\emph{pure} entangled state with $V(\rho_{pure})=V_{min}$. Our proof
does not rule out the possibility of ``entanglement purification'',
i.e., of transforming a large number of Gaussian mixed entangled
states into a (asymptotically) pure entangled state with the same
value of $V$ with Gaussian means. First calculations indicate that
this might indeed be possible. These results will be reported
elsewhere.
%%%%%%%%%%%%%%%%%%%%%%%%%%%%%%%%%%%%%%%%%%%%

\section{Gaussian positive maps}\label{Secposmaps}
In this Section we show how to extend the approach presented in
Section \ref{main} to include Gaussian \emph{positive} but not completely
positive (g-p) maps. 
To this end, we first define the set of Gaussian operators, generalizing
Gaussian density matrices to not self-adjoint operators. 
Every operator $A\in\cBH$ is completely determined by
$\chi_A(x):=\tr[A W(x)]$ \cite{Petz}.  
It follows that $A$ may be written in terms of $\chi_A$ as \cite{Perelomov}
\begin{equation}\label{densmat}
A = \pi^{-n} \int_{\RR^n}\!dx \chi_A(x) W(-x),
\end{equation}
We define  the set of Gaussian operators on $\cH$ by
\begin{equation}\label{AppGaussOp}
\cGH := \{A : A = \int_{\RR^{2n}}\!\!dx
e^{-\frac{1}{4}x^T\gamma x + ib^Tx-c}W(x)\}, 
\end{equation}
where $\gamma^T=\gamma\in M_{2n}(\CC), b\in\CC^{2n}$, and $c\in\CC$.
It is straight forward to check that $A$ is (1) \emph{bounded} iff
$\Re\gamma> 0$; (2) self-adjoint iff $\Im\gamma=0, \Im b=0, \Im c=0$;
and (3) \emph{positive} iff s.a. and $\gamma\geq iJ$; and (4)
$\tr(A)=\pi^ne^{-c}$, where $\tr[W(x)] = \pi^n\delta(x), x\in\RR^{2n}$ was
used.  To prove (3), consider one mode, $\gamma=g\id_2$. The
corresponding operator is $\forall g\geq0$ diagonal in the number
basis (for $g\geq 1$ it describes the well-known thermal states of a
field mode) and the eigenvalues $\bra{n}A_g\ket{n}$ are seen to be all
positive iff $g\geq 1$, otherwise odd numbers correspond to negative
eigenvalues. Finally, recall that all selfadjoint Gaussian operators
can be transformed into (a tensor product of Gaussian operators of)
that form by quasifree unitaries, performing the normal mode
decomposition of $\rho$ or, equivalently, the symplectic
diagonalization of $\gamma$, which concludes the proof.

Now we turn to linear maps on $\cBH$. Generalizing Section \ref{main}
we define a \emph{Gaussian map} as a linear map $\cG:\cBH\to\cB(\cH')$
that maps Gaussian operators to Gaussian operators, i.e., 
$\cG[\cGH]\subset{\cal Q}(\cH')$.
Again, we can use the isomorphism of \cite{isomorphism} to show that
all Gaussian maps on $\cBH$ correspond to Gaussian operators on
$\cBH\otimes\cBH$. i.e., they may be described by a matrix $\Gamma$, a
vector $D$, and a phase/normalization constant $C$. 
Then we can use \Eqref{GviaTelep} to calculate how the Gaussian map
$\cG$ corresponding to $(\Gamma,D,C)$ acts on the Weyl operator
$W(x)$. One finds
\[
\cG[W(x)] = \int dy
e^{-\frac{1}{4}{x\choose y}^T\tilde\Gamma {x\choose
y} +i\tilde D^T{x\choose y}- C}
W(y),
\]
where $\tilde\Gamma,\tilde D$ belong to the partial transpose of the operator
$G$ isomorphic to $\cG$.

One quickly convinces oneself that the map $\cG$ is selfadjoint, that
is  $\cG(A^\dagger)^\dagger = \cG(A)$ iff $G$ is selfadjoint, i.e.,
iff $\Im\Gamma=0, \Im D=0, \Im C =0$.  

A Gaussian map $\cG$ is called a \emph{Gaussian positive map} (g-p
map) iff it maps $\cGH_+$ to $\cGH_+$. 
In terms of matrices this means that $\cG$ is g-positive iff $\cG(\gamma)\geq
iJ\forall \gamma\geq iJ$. Expressing the action of $G$ through its matrix
$\Gamma$ this is equivalent to the condition
\[
\tilde\Gamma_1 -
\tilde\Gamma_{12}\frac{1}{\tilde\Gamma_2+\gamma}\tilde\Gamma_{12}^T\geq
iJ\hspace*{1.5ex}\forall \gamma\geq iJ.
\]
Inverted this inequality it is seen to be equivalent to  $\tilde\Gamma_1\geq
iJ$ and
\begin{equation}\label{gpm2}
\gamma + \underbrace{
\tilde\Gamma_2 -
\tilde\Gamma_{12}^T\frac{1}{\tilde\Gamma_1-iJ}\tilde\Gamma_{12}}_M\geq
0\hspace*{1.5ex}\forall \gamma\geq iJ,
\end{equation}
which can now be written in a $\gamma$-independent way as
\begin{equation}\label{posCrit}
\text{min}_{z\in\CC^{2n}}\text{max}\left\{
z^\dagger(M+iJ)z,z^\dagger(M-iJ)z \right\}\geq 0.
\end{equation}
To see that this is equivalent to Cond. (\ref{gpm2}) note that for any
$z=z_r+iz_i\in\CC^{2n}$ there exists a symplectic map $S$ such that
$z^\dagger SS^T z = |iz^\dagger Jz|=2|z_r^TJz_i|=c^2$; this can be
seen immediately by extending $\{s_1=z_r/c,s_2=z_i/c\}$ to a
symplectic basis $\{s_k\}$ and defining $S$ by $Ss_k=e_k$, where $e_k$
refers to the canonical basis. Therefore, if cond. (\ref{gpm2}) is to
hold for all $\gamma$, it holds in particular for $\gamma=S^TS$.

The minimum of these maxima can for given $\Gamma$ be efficiently
sought numerically, thus providing a practical characterization of all
positive Gaussian maps.  We emphasize, that such a practical
characterization of positive maps is not currently available for
general maps on $d$-level systems. 

Condition (\ref{posCrit}) says that the matrix $\Gamma$ of a positive map
can be such that neither $\tilde N:=\Gamma-iJ\geq0$ nor
$N:=\tilde\Gamma-iJ\geq0$, but there may be no vector $x\in\CC^{2n}$
such that both $x^T\tilde Nx$ and $x^TNx$ are negative. Examples are
(i)  cp maps ($\Gamma\geq iJ$, cf.\ Section
\ref{main}); (ii) maps for which $\Gamma\not\geq iJ$ but
$\tilde\Gamma\geq iJ$. These are \emph{decomposable} positive maps
(such as transposition); (iii) Gaussian maps for
which neither $N$ nor $\tilde N$ is positive can also be constructed;
these and the case of \emph{non-decomposable} g-p maps, i.e. those
that are not derived from transposition and the relation of
g-positivity to the usual notion of positivity will be discussed
elsewhere \cite{GPosMaps}.

%%%%%%%%%%%%%%%%%%%%%%%%%%%%%%%%%%%%%%%%%%%% 
\section{Conclusions}                                                                      
We have characterized all the physical actions that can be performed
using linear optics, squeezers, and homodyne measurements. We have
also characterized those that can be implemented with LOCC and those
that can be implemented using ppt-preserving maps. We have
used the methods developed in the preceding sections to show that
Gaussian states cannot be distilled by local Gaussian operations and
classical communication.  

Finally we have extended the definitions given before to general
linear maps that map Gaussian states to Gaussian states and provided a
complete characterization of \emph{positive} Gaussian maps. This
emphasizes that Gaussian states are worth studying not only because of
their experimental relevance (which will reduce as non-Gaussian states
become more accessible), but also on mathematical grounds that this
class of states is simple enough to derive strong results while being
large enough to encompass most (if not all) aspects of entanglement.
                                                                      
\appendix
\section{Properties of $V(\gamma)$}\label{V}
In this section we collect a number of useful properties of the
quantity $V(\gamma)$ introduced in Sec. \ref{Secbipart}b. 
$V$ is defined for bipartite CMs $\gamma$ (or, equivalently, for
Gaussian states $\rho_{\gamma,d}$) as 
\begin{equation}
V(\gamma) := \mathrm{max}_{\gamma_A,\gamma_B\geq
iJ}\left\{ p\leq1 : \gamma\geq p(\gamma_A\oplus\gamma_B)\right\}. 
\end{equation}

\noindent\textbf{(1) $V$ for more than one state: }
\begin{equation}
V(\gamma\oplus\gamma') = \mathrm{min}\left\{V(\gamma),V(\gamma')\right\},
\end{equation}
that is, $V$ does not decrease when several entangled states
are joined together. Rather, $V$ of the combined state is given by the
smallest $V$ of the individual states. 

To see this, let $v=V(\gamma), v'=V(\gamma')$. Clearly, 
$V(\gamma\oplus\gamma')\geq \mathrm{min}\{v,v'\}$ since by definition
of $V$ we have  $\gamma\oplus\gamma'\geq 
v(\gamma_A\oplus\gamma_B)\oplus  v'(\gamma_A'\oplus\gamma_B')\geq
\mathrm{min}\{v,v'\}(\gamma_A\oplus\gamma_B\oplus\gamma_A'\oplus\gamma_B')$.
On the other hand
$V(\gamma\oplus\gamma')\leq\mathrm{min}\{v,v'\}$ since 
$\gamma\oplus\gamma'\geq
V(\gamma\oplus\gamma')(\gamma_{AA'}\oplus\gamma_{BB'})$ also holds for
the reduced states with subsystems $AB$ or $A'B'$ traced out.
More generally, it follows that $V(\oplus_k\gamma_k) =
\mathrm{min}\{V(\gamma_k)\}$. \\

\noindent\textbf{(2) An upper bound for $V$: }
\begin{equation}
V(\rho_\gamma)\leq\mathrm{min}\{\lambda_{min}(\gamma),1\}, 
\end{equation}
where $\lambda_{min}(\gamma)$ is the 
smallest symplectic eigenvalue (smaller than 1) of the CM
$\tilde\gamma$ of the partially transposed state
$\rho_\gamma^{T_A}$ \cite{FN_neg}. $\lambda_{min}<1$ is necessary and
sufficient for the corresponding state to have a non-positive partial
transpose.\\  

\noindent\textbf{(3) $V$ for the maximally entangled state $\ket{\Phi}$: }
\begin{equation}
V(\Phi)= \lim_{r\to\infty}V(\gamma(r)) = 0,
\end{equation}
since $\lambda_{min}(\gamma(r))=e^{-r}$.\\

\noindent\textbf{(4) $V$ and negativity \cite{Vida02}: }
For $1\times N$ systems (i.e., in systems where no 
ppt-entanglement exists \cite{GBE}) we also have
$V(\rho)\geq\lambda_{min}$ since in that case
$\gamma/\lambda_{min}$ has positive partial transpose and therefore
is separable. This shows that $V(\rho)$ 
is related to the negativity measure of entanglement
\cite{Vida02}, and for $1\times N$ systems $-\log_2[V(\rho_\gamma)]$
coincides with the log-negativity (up to a factor). In contrast, for
ppt-entangled Gaussian states \cite{GBE} $V(\gamma)$ is strictly smaller
than 1, while the negativity of such states is zero
($\lambda_{min}\geq 1$). \\

\noindent\textbf{(5) $V$ as a Gaussian measure of entanglement: }
We have seen that $V(\gamma)$ does not decrease under local
Gaussian operations. Hence it can be considered a measure of
entanglement for Gaussian states. In view of (4), we see that $V$ does
quantify both npt and ppt Gaussian entanglement -- in contrast to most
other measures of entanglement calculated to date. \\

\noindent\textbf{(6) $V$ is computable: }
It is worth pointing 
out that $V(\gamma)$ is also \emph{computable}, as one can use the
separability criterion derived in \cite{GSepCrit} to find the largest
$p$ for which $\gamma/p$ is separable.\\

\emph{Note added: } Upon completion of this work we learned that
Jarom\'{\i}r Fiur\'a\v sek \cite{Fiurasek} independently arrived at a
similar description of general Gaussian operations and, in particular,
of Gaussian LOCCs. 
                                                                      
\begin{acknowledgments}                                               
We acknowledge very valuable discussions with Jens Eisert. GG also
thanks Jarom\'{\i}r Fiur\'a\v sek for interesting and useful
discussions. We acknowledge support by the European Union under the
project EQUIP (contract IST-1999-11053).
\end{acknowledgments}                                                 
                                                                      
%%%                                                                   
  

\begin{thebibliography}{99}
\bibitem{footnote}                                                    
Although this maximally entangled state does not belong to the        
Hilbert space, it can always be considered as a limit of a proper     
pure state.
\bibitem{FN_improper} 
A mathematically correct way of dealing with the ``projections into
improper eigenstates of $X$'' would be via a limit procedure: define
$\ket{\Phi_r}=\cosh(r)^{-2}\sum_{k\geq0}\tanh(r)^k\ket{kk}$ and a
``Bell-POVM'' $\left\{ W(x)\ket{\Phi_r}\bra{\Phi_r}W(-x):x\in\RR
\right\}$.  Then we define $E_{12}^{(r)} =
(\cE\otimes\id)(\ket{\Phi_r}\bra{\Phi_r})$ and finally $\cE(\rho) =
\lim_{r\to\infty}\tr_2[(E_{12}^{(r)})^{T_2}\rho_2]$. The limit can be
performed directly on the level of correlation matrices yielding the
expressions \Eqsref{gammap_gamma}.
\bibitem{useisomorphG}
The correspondence of Gaussian operations and Gaussian states has been
used in \cite{FiurasekCVTel} to optimize continuous variable
teleportation with Gaussian means.
\bibitem{FN_neg} Here we use that for two strictly positive matrices
$A\geq B$ implies that $\lambda_k(A)\geq\lambda_k(B)$, where the
$\lambda_k$ are the (descendingly ordered) symplectic eigenvalues of
the respective matrix.
\end{thebibliography}
\end{document}